\documentclass[twoside,fleqn]{article}
\usepackage{espcrc2}

\def\gsim{{\mathrel{\raise2pt\hbox to 8pt{\raise -5pt\hbox{$\sim$}\hss{$>$}}}}}
\def\rsim{{\mathrel{\raise2pt\hbox to 8pt{\raise -5pt\hbox{$\sim$}\hss{$>$}}}}}
\def\lsim{{\mathrel{\raise2pt\hbox to 8pt{\raise -5pt\hbox{$\sim$}\hss{$<$}}}}}

\def\half{{\textstyle{1\over2}}}
\def\CO{\mathcal{O}}
\def\CS{\mathcal{S}}

\def\CV{\mathcal{V}}
\def\hat{\widehat}
\def\tr{\textrm{tr}}
\def\Tr{\textrm{Tr}}

\begin{document}

\title{
       \begin{flushright}\normalsize
	    \vskip -1.4 cm
            LA-UR-03-6623, SNUTP-03-019, UW-PT 03-12
       \end{flushright}
Improved bilinears in unquenched lattice QCD
\thanks{
Research supported in part by US-DOE grant
KA-04-01010-E161 and
contract DE-FG03-96ER40956/A006, and by
BK21, the SNU foundation \& Overhead Research fund and
KRF contract KRF-2002-003-C00033.}
}

\author{Tanmoy Bhattacharya$\rm ^{a}$,
Rajan Gupta\address{Theoretical Division, Los Alamos National Lab, Los Alamos,
         New Mexico 87545, USA},
Weonjong Lee\address{School of Physics, Seoul National University, Seoul, 151-747, Korea},
Stephen R. Sharpe$\rm ^{c}$ and 
Jackson~M.~S. Wu\address{Physics Department,
University of Washington, Seattle, WA 98195-1560, USA}}
      
\begin{abstract}
We summarize the extent to which one can use Ward identities (WI) to 
non-perturbatively improve
flavor singlet and non-singlet bilinears with
three flavors of non-degenerate dynamical Wilson-like fermions.
\vspace{-0.2in}
\end{abstract}

\maketitle


It is important to pursue simulations with 
Wilson-like fermions, despite their greater computational
cost than staggered fermions, because one avoids the
theoretical uncertainty associated with
taking roots of the fermion determinant.
One drawback with Wilson fermions is that they lead to
discretization errors of $O(a)$, slowing the approach to
the continuum. This can be ameliorated using an $O(a)$
improved action, with the relevant improvement
coefficient, $c_{SW}$, determined non-perturbatively using the
method of Ref.~\cite{alpha}.
One also wants to improve the matrix elements of operators,
and a technique for doing so has been worked out for non-singlet
quark bilinears with degenerate dynamical quarks~\cite{alpha}.

Here we sketch the generalization of this improvement
program to the realistic case of three non-degenerate dynamical
quarks, and to all bilinears, including flavor singlets.
The inclusion of flavor singlets is a necessary part of
the technique, and is furthermore of phenomenological interest.

This study is the completion of the work presented
in \cite{lat99}, which was incomplete and partly
incorrect. Many details must necessarily be skipped in
this short writeup; a complete discussion,
including the generalization
to two and four flavors, will be presented in \cite{UQWI}.

We assume that the following steps in the improvement
program have already been taken: (1) $c_{SW}$ and $c_A$ have been determined
by enforcing the PCAC relation for non-singlet axial 
currents with degenerate quarks \cite{alpha};
(2) $\kappa_c$ is known from the
vanishing of the ``Ward identity mass'', so that the bare
quark masses $a m_j = 1/2 \kappa_j - 1/2\kappa_c$ can be defined;
(3) Non-singlet vector and tensor bilinears have been improved
by the addition of the standard $c_V$ and $c_T$ 
terms~\cite{alpha}, which are then determined by enforcing axial WI
in the chiral limit~\cite{GuagnelliSommer,WIPLB,Martinelli}---so that
all improved non-singlet bilinears $O^{(jk),I}$ are known in the
chiral limit;\footnote{%
Notation: $O$ is a $3\times3$ flavor matrix, 
flavor labels $i,j,k$ are not summed, 
and are implicitly taken to be distinct.}
%
and (4) $b_g$ has been determined,
using the methods of Refs.~\cite{alpha,Martinelli} or 
that presented below, and
the bare coupling $g_0$ is adjusted, as the mass matrix
$M=\mathrm{diag}(m_1,m_2,m_3)$ is varied, so that the effective
coupling $g_0^2 (1 + a\, b_g \tr M /N_f)$ is held fixed.


As already noted in \cite{lat99}, the generalization
to non-degenerate quarks introduces many new improvement
coefficients. The improved, renormalized off-diagonal bilinears
($O=S,P,V,A,T$),
\[
\widehat\CO^{(jk)} \! =\! Z_O
\big[1\! +\! a \bar b_O \tr M 
\!+\! \ a b_O \half (m_j+m_k)\big] 
\CO^{(jk),I}
\,,
\]
require one more coefficient, $\bar b_O$, than
in the quenched case, while the diagonal non-singlet bilinears
require another new coefficient, $f_O$:
\begin{eqnarray*}
\lefteqn{
\widehat\CO^{(jj)} \!-\! \widehat\CO^{(kk)} \!=\! 
Z_O \bigg[ (1 \!+\! a \bar b_O \tr M) (\CO^{(jj),I} \!\!-\! \CO^{(kk),I})}  
\nonumber\\
&& \!\!\!\!\!\!\!\!\!\!\!\!\!+
a b_O (m_j \CO^{(jj)} \!-\! m_k \CO^{(kk)}) \!+\! a f_O (m_j\!-\!m_k) \tr\CO
\bigg]\,.
\end{eqnarray*}
For $O=S$ there is also mixing
with the identity operator, missed in~\cite{lat99}, and
discussed below.

Flavor singlet bilinears require new coefficients.
In the chiral limit one has 
\begin{eqnarray*}
(\tr A_\mu)^I &=& \tr A_\mu + a\, \bar c_A \partial_\mu \tr P 
\\
(\tr V_\mu)^I &=& \tr V_\mu + a\, \bar c_V \partial_\nu \tr T_{\mu\nu}
\\ 
(\tr T_{\mu\nu})^I &=& \tr T_{\mu\nu} + a\, \bar c_T \left[
\partial_\mu \tr V_{\nu} - \partial_\nu \tr V_{\mu} \right]
\\
(\tr P)^I &=& \tr P + a\, g_P \Tr(F_{\mu\nu} \tilde F_{\mu\nu}) \\
(\tr S)^I &=& a^{-3} e_S
+ \tr S + a\, g_S \Tr(F_{\mu\nu} F_{\mu\nu}) \,.
\end{eqnarray*}
Note that gluonic operators are
needed to improve $\tr P$ and $\tr S$ 
(but not $\tr T$, as erroneously concluded in~\cite{lat99}),
and the need to subtract the identity operator
from $\tr S$.
Away from the chiral limit one has
\[
\widehat{\tr\CO} = Z_O r_O \left[
(1 \!+\! a\bar d_O \tr M) \tr(\CO)^I \!+\! a d_O \tr(M \CO) \right]
\,, \nonumber
\]
with new coefficients $r_O$, $d_O$ and $\bar d_O$.
$r_A$ is scale dependent since the singlet axial current has anomalous
dimension, while the other $r_O$ are scale independent.
For $O=S$ we have implicitly subtracted mass dependent mixing
with the identity operator, as discussed below.

Finally, the improved quark masses are:
\begin{eqnarray*}
\widehat{\tr\lambda M} &\!\!=\!\!& Z_m \left[
(1 \!+\! a\bar b_m \tr M) \tr\lambda M \!+\! a b_m \tr(\lambda M^2)
\right]
\\
\widehat{\tr M} &\!\!=\!\!& Z_m r_m \left[
(1 \!+\! a\bar d_m \tr M) \tr M \!+\! a d_m \tr(M^2) \right]
\end{eqnarray*}
where $\lambda$ are Gell-Mann matrices.
$Z_m$ is scale dependent, while $r_m$ is not.
Note that there is no independent $f_m$-like term.


The coefficient $e_S$ would need to be determined
to an accuracy of $a^4$ to improve $\tr S$, which 
appears impractical.
However, to calculate hadronic matrix elements {\em one must subtract
disconnected contributions anyway}, and this completely cancels 
the $e_S$ term. Similarly, the
 $e_S$ contribution can be canceled when implementing
WI by subtracting disconnected contributions.
In this way one can avoid the problem, except when calculating
the vacuum expectation value. The same issue arises for the
diagonal non-singlet scalars away from the chiral limit,
and has the same resolution.


The first step in determining the improvement coefficients is standard:
one enforces the
normalization of vector charges for a selection of hadrons 
for various quark masses.
This determines $Z_V$, $r_V$, $b_V$, $\bar b_V$, $f_V$, $d_V$ and
$\bar d_V$, i.e. all the vector coefficients except $c_V$ and $\bar c_V$.
The latter do not contribute to the charges.

In Ref.~\cite{lat99} it was claimed that one can learn about
other improvement
coefficients by enforcing the vector transformation properties
of bilinears. This is incorrect. The essential reason is that
the lattice theory does not break the vector symmetry. The relations
that were to be used can be shown to be automatically satisfied.


The next step is to relate the coefficients for scalar bilinears
and the quark mass, enforcing the standard continuum
convention $Z_S Z_m=1$. This is unchanged from Ref.~\cite{lat99}---the
new $e_S$ term does not contribute. The result is that
all other coefficients for the scalar can be expressed
in terms of those for the quark mass, so that the former are
redundant. In particular, we find
$g_S = b_g/(2 g_0^2)$,
and the constraint
$d_S = b_S + 3 \bar b_S$, which follows from the
absence of an $f_m$ term.



We next consider ``two-point'' axial WI, i.e. 
enforce the axial divergence relations on improved quantities
while varying quark masses. In addition to the off-diagonal
currents used in \cite{alpha,Petronzio} and \cite{lat99}, 
we now use flavor diagonal currents:
\begin{eqnarray*}
\lefteqn{\langle \partial_\mu (\hat A_\mu^{(jj)}(x)\!-\! 
\hat A_\mu^{(kk)}(x)) J(0) \rangle = O(a^2) +} 
\nonumber\\
&&\!\!\! 2 \left\langle \left[\hat m_j \hat
P^{(jj)}(x) \!-\! \hat m_k \hat P^{(kk)}(x)\right] J(0)\right\rangle
\ \  [x\ne 0]\,.
\end{eqnarray*}
The RHS involves singlet parts of $M$ and $P$, illustrating how
these cannot be avoided. When expanded out in terms
of bare lattice operators and masses, there are several independent operators
on both sides, as well as quark-connected and disconnected contractions,
so this relation will be complicated to implement.
Nevertheless, it is a powerful constraint, yielding eight
independent 
combinations of coefficients~\cite{UQWI}.


The final step is to enforce
the axial transformation properties of the bilinears---the
so-called ``three-point'' axial WI.
Relations of the form
\[
\delta_A^{(ij)} \widehat{O}^{(jk)} = \widehat{\delta \CO}
{}^{(ik)}
\]
are the simplest to implement as they involve 
no quark-disconnected contractions.
They have been used extensively in quenched simulations~\cite{alpha,WIPLB},
and can be used in the chiral limit to determine
$c_V$ and $c_T$ as noted above.
The generalization to the unquenched case is straightforward.


To determine many of the new coefficients, however, 
one must use other three-point WI. 
Consider first
the transformation properties of the singlet axial current:
\begin{eqnarray*}
O(a^2) &\!\!=\!\!&\langle \delta^{(ij)}_{A}\CS\ \tr[\widehat{A_\mu}(y)]\ 
				J^{(ji)}(z) \rangle  \\ 
\delta^{(ij)}_{A}\CS &\!\!=\!\!& a^4 \sum_{\CV}
 \Big[ (\widehat{m}_{i} + \widehat{m}_{j})
\widehat{P}^{(ij)} - \partial_{\mu} \widehat{A}^{(ij)}_{\mu} \Big] \,.
\end{eqnarray*}
Here the four-volume $\CV$ contains $y$ but not $z$.
Contact terms are avoided by setting $\hat m_i\!+\!\hat m_j=0$.
The above relation allows the determination of
the {\em relative size} of different terms in $\tr\widehat A$,
though not the overall normalization.
We find that one can obtain
$\bar c_A$ and $d_A$.
Replacing $\tr A$ with $\tr V$ one can similarly obtain $\bar c_V$
and $d_V$.


Further coefficients result from enforcing
\[
 \delta_A^{(ij)}\tr[\widehat{T_{\mu\nu}}]
  = -\epsilon_{\mu\nu\rho\sigma} 
 \widehat{T_{\rho\sigma}} {}^{(ij)}\,.
\]
This determines the
relative size of different terms in $\tr\widehat T$
(giving $\bar c_T$ and $d_T$)
and the ratio of normalizations of LHS and RHS
(giving $r_T$ and $\bar b_T-\bar d_T$).

%
%


The scalar case is of particular interest, i.e.
\[
\langle \delta^{(ij)}_{A}\CS\ \tr[\widehat{S}(y)]\ 
				J^{(ji)}(z) \rangle \!=\!
2 \langle \hat{P}^{(ij)}(y) J^{(ji)}(z) \rangle
\]
From this it appears that one can determine $g_S$ and $d_S$ from
the relative normalizations on the LHS, and $r_S$ and
$\bar b_P-\bar d_S$ from the ratios of normalizations
of the LHS and RHS.
There is, however, a subtlety because of the identity
operator contribution to $\tr S$.
Formally, this contribution vanishes since it
it is invariant under axial transformations, and
in particular
\[
\langle \delta^{(ij)}_{A}\CS\ 	J^{(ji)}(z) \rangle = O(a^2)
\,.
\]
However, the divergent coefficient $\propto 1/a^3$ invalidates
this argument for the lattice calculation,
so that one must explicitly subtract 
the disconnected contribution
\[
\langle \tr[\widehat{S}(y)]\rangle \times 
\langle \delta^{(ij)}_{A}\CS\ J^{(ji)}(z) \rangle 
\]
from the LHS to remove the $e_S$ term.

This provides a new method for determining 
$b_g=2 g_0^2 g_S$. Note that
one can carry out this determination in the chiral limit,
so that one does not need to know $b_g$ {\em a priori}.

The final WI giving new information is\footnote{%
This was not considered in \cite{lat99}.}
\vspace{-0.05truein}
\[
\delta_A^{(jk)}\widehat{T_{\mu\nu}}{}^{(kj)}(y)
\!=\! -\half\epsilon_{\mu\nu\rho\sigma} 
\bigg[
\hat{T_{\rho\sigma}}^{(jj)}(y)\!+\!\hat{T_{\rho\sigma}}^{(kk)}(y)
\bigg]
\]
\vspace{-0.05truein}
The RHS contains singlet and non-singlet pieces, 
and so gives new information
on improvement coefficients, although
at the cost of requiring quark-disconnected contractions.
We find that, in addition to constants determined previously,
this WI allows the determination of $f_T$.
A new subtlety here is that the contact terms to be
avoided cannot be obtained simply using the prescription for
 equation-of-motion
operators given in~\cite{WIPLB}. 

Combining results from all the WI considered above,
we find that one can determine all the renormalization
and improvement coefficients introduced above except
(1)
the scale-dependent constants $Z_T$, $Z_S Z_P$, and $r_A$
(whose determination requires use of a method like NPR, which
fixes operator normalizations by some convention);
and (2)
the two scale independent quantities $\bar d_A$ and 
$\bar b_T+\bar d_T$.\footnote{%
This corrects the conclusion of \cite{lat99} that there
were three undetermined scale independent quantities.}

Several issues merit further investigation.
For example,
is it possible to determine $\bar d_A$ and $\bar b_T+\bar d_T$
using other Ward identities?
How do the results generalize to two and four flavors
(see Ref.~\cite{UQWI})?
What is lost if one works at fixed $g_0^2$, which is
easier in practice?
And to what extent does the proposal of Ref.~\cite{FrezzottiRossi}
(which requires an even number of flavors in its simplest
form) remove the need to determine the improvement coefficients?



\begin{thebibliography}{9}
\bibitem{alpha}
K.Jansen {\it et al.},
Phys.Lett.B {\bf 372}, 275(1996)
[arXiv:hep-lat/9512009];
M.L\"uscher~{\it et~al.},
Nucl. Phys. B {\bf 478}, 365 (1996) [arXiv:hep-lat/9605038].


\bibitem{lat99}  
T.Bhattacharya{\it~et~al.},~Nucl. Phys. B
(Proc. Suppl.)
{\bf 83-84},902(2000) [arXiv:hep-lat/9909092].

\bibitem{UQWI}
T.Bhattacharya {\it et al.},
in preparation.

\bibitem{GuagnelliSommer} 
M.Guagnelli~\&~R.Sommer,~Nucl.Phys.B(Proc. 
Suppl.){\bf 63},886(1998)~[arXiv:hep-lat/9709088]

\bibitem{WIPLB} 
T.Bhattacharya{\em~et~al.},Phys.Lett.B{\bf 461},79
(1999) [arXiv:hep-lat/9904011];
Phys.Rev.D {\bf 63}, 074505(2001)
[arXiv:hep-lat/0009038].

\bibitem{Martinelli}
G.Martinelli {\em et al.},
Phys. Lett. B {\bf 411}, 141(1997)
[arXiv:hep-lat/9705018].

\bibitem{Petronzio}
G.M.deDivitiis and R.Petronzio,
Phys.Lett.B {\bf 419}, 311(1998)
[arXiv:hep-lat/9710071].

\bibitem{FrezzottiRossi}
R.Frezzotti and G.C.Rossi, arXiv:hep-lat/0306014.


\end{thebibliography}
\end{document}